\documentclass[12pt, a4paper]{article}

\usepackage[final]{showkeys}
\usepackage{a4}
\usepackage{amsmath}
\usepackage{amssymb}
\usepackage{epsfig}

\newcommand{\imag}{\textrm{i}}

\begin{document}

\title{Damped collective motion of many body systems: \\
       A variational approach to the quantal decay rate}

\author{Christian~Rummel and Helmut Hofmann \\
\small\it{Physik-Department der Technischen Universit\"at M\"unchen,
          D-85747 Garching, Germany}}

\date{}

\maketitle

\begin{abstract}
We address the problem of collective motion across a barrier like
encountered in fission. A formula for the quantal decay rate is
derived which bases on a recently developed variational approach
for functional integrals. This formula can be applied to low
temperatures that have not been accessible within the former PSPA
type approach. To account for damping of collective motion one
particle Green functions are dressed with appropriate
self-energies.
\end{abstract}

\section{Introduction}
\label{intro}

For a particle moving in a metastable potential  J.S.~Langer
\cite{langer} has shown that the decay rate
\begin{equation}\label{rate-langer}
R = \frac{|\kappa|}{\pi T} \ \textrm{Im} \,\mathcal{F}
\end{equation}
can be calculated from the partition function $\mathcal{Z}$
via the imaginary part of the free energy
\begin{equation}\label{FfromZ}
\mathcal{F} = -T \ \textrm{ln} \,\mathcal{Z} \,.
\end{equation}
(As usual temperature is measured in MeV such that the Boltzmann
constant is put equal to $k_{\textrm{B}} \equiv 1$.) The parameter
$1/\kappa$ (where $\kappa < 0$) sets the time scale for the
solution $\langle q \rangle_{t} = \langle q \rangle_{t=0} \ \exp
(-\kappa t)$ which describes the motion of the average trajectory
away from the barrier top. Using path integrals Langer's formula
has extensively been exploited in the framework of the
Caldeira-Leggett model (CLM) \cite{caa.lea:ap:83} for dissipative
quantum systems. Within this model the decay rate has been studied
in great detail from high to very low temperatures. The region of
the ``crossover temperature'' $T_{0}$, below which quantum
tunneling becomes more important than pure thermal activation has
been treated in \cite{larkin,grabert,grh.olp.weu:prb:87}; for
reviews on this subject see e.g. \cite{hap.tap.bom:rmp:90,weissu}.

In the present paper we want to adapt similar methods to systems
like finite nuclei, where the dynamics of the particles is
governed by a varying mean field. Different to the CLM this field,
and hence the "heat bath", changes with the variable which
parameterizes the path along which the free energy exhibits the
barrier. A first attempt along these lines was reported in
\cite{ruc.hoh:pre:01}. There, the partition function was
calculated by accounting for correlations around the mean
trajectory which were of  RPA-type. In different context such an
approach to $\mathcal{Z}$ has been phrased "Perturbed Static Path
Approximation" (PSPA), for details see
\cite{pug.bop.brr:ap:91,ath.aly:npa:97,rossignoli}. As this method
is based on a quadratic expansion of the effective action about
stationary points it suffers the same breakdown as an analogous
expansion within the CLM; the rate diverges at the $T_0$. In the
present paper we like to overcome this problem by employing a
variational approach to the quantum partition function, suggested
recently in \cite{rummel:phd,ruc.hoh:var-PF}.

At finite temperature nuclear collective motion is damped. In
\cite{ruc.hoh:pre:01} this feature was simulated by applying
simple energy smearing to the results obtained within the
independent particle model (IPM). In the present paper we like to
apply a more microscopic picture. The latter bases on the
replacement of the one-body Green function of the IPM by dressed
ones, accounting in this way for incoherent scattering processes
among the particles. In this way connection is established to
transport theory for damped collective motion (see e.g.
\cite{siemensetal:84,hoh:pr:97}). Indeed, by using linear response
functions we are able to introduce microscopically
transport coefficients like the inertia $M$, friction $\gamma$ or
the stiffness $C$, which then may serve as input for the
calculation of the decay rate of metastable systems.

The paper is organized as follows: In Sect.~\ref{sec-partfunc} we
briefly review several approximations to the partition function
$\mathcal{Z}$ of interacting many body systems. Thereafter in
Sect.~\ref{chap-diss} we take into account residual interactions,
apply linear response theory to treat damped harmonic motion and
establish the connection to microscopic transport theories. In
Sect.~\ref{sec-modify} we modify the methods of
Sect.~\ref{sec-partfunc} such that dissipation can be treated.
After the derivation of formulas for the decay rate of metastable
interacting many body systems in Sect.~\ref{decay} we want to
estimate the accuracy of our results. In the ideal case this
should be done with an exactly solvable model for the many body
system. However, to the best of our knowledge for damped
collective motion, the case we are interested in, such a model
does not exist. For this reason so far we are only able to test in
Sect.~\ref{reliability} our approach at the example of a particle
moving in a one-dimensional potential under the influence of
damping, leaving a more involved analysis for many body systems
for future work. Finally we like to close by discussing our
findings and drawing some conclusions.

\section{Approximations to the partition function}
\label{sec-partfunc}

The collective degree of freedom shall be introduced through a mean
field approximation to a separable interaction of the type
\cite{bohra.mottelsonb.2} $k\hat{F}^2/2$.
It is to be added to the unperturbed Hamiltonian
$\hat{H}$, which at first shall be assumed to be of one body
nature. The total $\hat{\cal H}$ thus is given by
\begin{equation}\label{twobodham}
\hat{\cal H} =
\hat{H} + \frac{k}{2} \,\hat{F}\hat{F} \,.
\end{equation}
For $k < 0$ the interaction is attractive and leads to
iso-scalar modes in the nuclear case. The operator $\hat F$
is meant to represent an effective generator of
collective motion along a fission path. For the evaluation of the
partition function $\mathcal{Z}$ we want to exploit functional
integrals in imaginary time $\tau = 0 \ldots \hbar\beta$ with
$\beta = 1/T$. The two body interaction may then be treated by the
Hubbard-Stratonovich transformation (HST) to the mean field
Hamiltonian
\begin{equation}\label{1bHam}
\hat{\mathcal{H}}_{\textrm{HST}}[q(\tau)] =
\hat{H} + q(\tau) \,\hat{F} \,.
\end{equation}
Here, the $q(\tau)$ represents the collective variable.
Following \cite{feynmanr.hibbsa} it will be treated by an expansion
around an average value $q_{0}$ defined through
$\hbar\beta q_0 =\int_{0}^{\hbar\beta} d\tau \,q(\tau)$.
The deviation will be expanded into the Fourier series
$q(\tau) - q_{0} = \sum_{r \ne 0} q_{r} \exp (\imag\nu_{r}\tau)$,
with the Matsubara frequencies $\nu_{r} = (2\pi/\hbar\beta) \,r$
and $q_{-r} = q_{r}^{*}$. As described elsewhere (see
e.g. \cite{ath.aly:npa:97,ruc.anj:epjb:02,ruc.hoh:var-PF}) the partition
function can be expressed by the following (ordinary) integral over the
$q_0$:
\begin{equation}\label{Z-athaly}
\mathcal{Z}(\beta) =
\sqrt{\frac{\beta}{2\pi |k|}}
\int_{-\infty}^{+\infty} \!\! dq_{0} \ \exp (-\beta
\mathcal{F}^{\textrm{SPA}}(\beta, q_{0}))
\ \zeta(\beta, q_{0})
\end{equation}
In this expression $\mathcal{F}^{\textrm{SPA}}$ is the
$q_{0}$-dependent effective free energy in the quasi-static picture,
left over here in the classical limit, where all $\tau$-dependent
deviations from $q_{0}$ can be neglected.
For convenience we introduce the effective free energy
\begin{equation}\label{Feff}
\mathcal{F}^{\textrm{eff}}(\beta, q_{0}) =
\mathcal{F}^{\textrm{SPA}}(\beta, q_{0}) -
\frac{1}{\beta} \ \textrm{ln} \,\zeta(\beta, q_{0}) \,,
\end{equation}
which includes quantum effects through the factor $\zeta$.
This correction still is a functional integral
in $q_{r}$-space and specified through an Euclidean action
$s_{\textrm{E}}(\beta,q_{0})$. Various existing approximations consist
in the way this action is treated. As already indicated through the
notation employed in (\ref{Z-athaly}) and (\ref{Feff}),
within the Static Path Approximation (SPA) the $\zeta^{\textrm{SPA}}$
is put equal to unity, which implies to discard all
fluctuations around $q_{0}$. Corrections to this approximation
(see e.g. \cite{ruc.hoh:var-PF})
are obtained by expanding systematically the action
\begin{equation}\label{defA}
s_{\textrm{E}} =
\frac{\hbar\beta}{|k|} \left(
\sum_{r,s \ne 0} \lambda_{rs} \,q_{r} q_{s}
+ \sum_{r,s,t \ne 0} \rho_{rst} \,q_{r} q_{s} q_{t}
+ \sum_{r,s,t,u \ne 0} \sigma_{rstu} \,q_{r} q_{s} q_{t} q_{u} \right)
+ {\cal O}(q_{r}^{5})
\end{equation}
around the static path $q_{0}$ to some power in the $q_{r}$, exhibited
here to fourth order.

Within the SPA+RPA \cite{pug.bop.brr:ap:91}, the Perturbed Static Path
Approximation (PSPA) \cite{ath.aly:npa:97} or the Correlated Static Path
Approximation (CSPA) \cite{rossignoli} the expansion
(\ref{defA}) is truncated after the second order
\begin{equation}\label{s-PSPA}
s_{\textrm{E}}^{\textrm{PSPA}} =
\frac{\hbar\beta}{|k|}
\sum_{r,s \ne 0} \lambda_{rs} \,q_{r} q_{s} \,.
\end{equation}
This approximation takes into account quantum fluctuations
around $q_{0}$ on the level of local RPA modes.
It turns out (see e.g. \cite{ruc.hoh:pre:01})
that the second order coefficients of the PSPA are
diagonal $\lambda_{rs} = \lambda_{r}/2 \ \delta_{r,-s}$,
with the $\lambda_{r}$ being given by
\begin{equation}\label{lambda}
\lambda_{r}(\beta,q_{0}) =
1 + k\,\chi(\imag\nu_{r}) =
\frac{\prod_{\mu} (\nu_{r}^{2} + \varpi_{\mu}^{2}(\beta,q_{0}))}
     {\prod_{k>l}' (\nu_{r}^{2} + \omega_{kl}^{2}(q_{0}))} \,.
\end{equation}
Here, the response function $\chi(\omega)$ has been introduced,
which is defined through the relation
\begin{equation}\label{defchi}
\delta\langle\hat{F}\rangle_{\omega} =
-\chi(\omega) \,\delta q(\omega)
\end{equation}
where $\delta q(\omega)$ means a general time-dependent deviation
from $q_{0}$. Mind that the averages have to be calculated from the
Hamiltonian (\ref{1bHam}) at {\em fixed} $q_{0}$.
In this way the $\chi(\omega)$ depends on $\beta$ and $q_{0}$.
Since this $\chi(\omega)$ only contains excitations of nucleonic nature,
\begin{equation}\label{nuclfreq}
\hbar\omega_{kl} = \epsilon_{k} - \epsilon_{l} \,,
\end{equation}
which are to be distinguished from collective ones
(to which we will come to below), we may call it the
{\em nucleonic} or {\em intrinsic} response function. In the following
the single particle energies shall be measured with respect to the
chemical potential $\mu$, which means to write
$\hbar\omega_{k}(q_{0}) = \epsilon_{k}(q_{0}) - \mu(q_{0})$.
The local RPA frequencies $\varpi_{\mu}(\beta,q_{0})$ appearing in
(\ref{lambda}) must be calculated from the secular equation
\begin{equation}\label{secular}
1 + k \chi(\varpi_{\mu}) = 0 \,.
\end{equation}
Note that the important factors in the product (\ref{lambda}) are
those which deviate from one. This happens whenever the
$\varpi_{\mu}^{2}$ is sizably different from the corresponding
$\omega_{kl}^{2}$. This is so in particular for the typical
collective modes, for which for stable isoscalar modes, for instance,
one has $\varpi_{\mu}^{2}\ll \omega_{kl}^{2}$.

For proper convergence of the path integral for
$\zeta^{\textrm{PSPA}}$ one needs the condition
(see e.g. \cite{ruc.anj:epjb:02,ruc.hoh:var-PF})
\begin{equation}\label{convcond}
\lambda_{1}(\beta,q_{0}) > 0 \,.
\end{equation}
It ensures the PSPA to be well defined such that the quantum
corrections can be written in the form
\begin{equation}\label{defCPSPA}
\zeta^{\textrm{PSPA}}(\beta, q_{0}) =
\prod_{r>0} \frac{1}{\lambda_{r}(\beta,q_{0})} \,.
\end{equation}
The restriction (\ref{convcond}) can easily be translated into a
condition on that temperature at which an unphysical divergence of the
quantum fluctuations around $q_{0}$ occurs: $T > T_{0}$. Here
\begin{equation}\label{T0}
T_{0} = \textrm{Max}_{q_{0}}
\,\frac{\hbar |\varpi_{\textrm{inst}}(q_{0})|}{2\pi}
\end{equation}
can be calculated from the {\em unstable} local RPA mode for which
$\varpi_{\textrm{inst}}^{2} < 0$. The latter is present whenever
$\mathcal{F}^{\textrm{SPA}}(q_{0})$ develops a barrier.

Obviously, the unphysical divergence at $T_{0}$ is a deficiency of
the PSPA. It can be cured by accounting for inharmonicities in the
expansion (\ref{defA}) of the action.
To be able to lower the breakdown temperature to $T_{0}/2$
it is sufficient to concentrate on the three coefficients
$\rho_{1,1,-2}$, $\rho_{-1,-1,2}$ and $\sigma_{1,1,-1,-1}$ and neglect
the remaining ones \cite{ruc.anj:epjb:02}. This improved approximation
was called extended Perturbed Static Path Approximation (ePSPA).
For $T > T_{0}/2$ the quantum correction factor may then be
expressed by the form
\begin{equation}\label{C-ePSPA}
\zeta^{\textrm{ePSPA}}(\beta,q_{0}) = \sqrt{\pi} \
\frac{x(\beta,q_{0})}{\lambda_{1}(\beta,q_{0})} \ \exp \left[
x^{2}(\beta,q_{0}) \right] \,\textrm{erfc} \left[ x(\beta,q_{0}) \right]
\ \prod_{r>1} \frac{1}{\lambda_{r}(\beta,q_{0})} \,.
\end{equation}
The quantity
\begin{equation}\label{defxB}
x = \sqrt{\frac{\beta}{4|k|B}} \ \lambda_{1} \qquad
\textrm{with} \qquad B = 6 \sigma_{1,1,-1,-1} - \frac{9 \rho_{-1,-1,2}
\rho_{1,1,-2}}{\lambda_{2}}
\end{equation}
is given in terms of the third and fourth order inharmonic terms
of the Euclidean action (\ref{defA}). The coefficients $\rho$ and
$\sigma$ can be expressed through one body Green's functions
\begin{equation}\label{Green-freq-IPM}
g_{k}^{(0)}(z) =
\frac{1}{\hbar} \,\frac{1}{z - \omega_{k}}
\end{equation}
corresponding to the static part of the Hamiltonian (\ref{1bHam}).
Details of the calculation can be found in
\cite{ruc.anj:epjb:02,rummel:phd}. With Fermi occupation numbers denoted
by $n(\epsilon_{k})$ the general coefficient $\sigma_{rstu}$
consists of a sum of terms like
\begin{eqnarray}\label{Cauchy}
& & \frac{|k|}{4!} \sum_{i,k,m,o} F_{io} F_{ki} F_{mk} F_{om} \times \\
& & \left\{ n(\epsilon_{i})
\ g_{o}^{(0)}(\omega_{i} + \imag\nu_{r})
\ g_{k}^{(0)}(\omega_{i} - \imag\nu_{s})
\ g_{m}^{(0)}(\omega_{i} - \imag\nu_{s+t})
\right. \nonumber \\
& & + \ n(\epsilon_{o})
\ g_{i}^{(0)}(\omega_{o} - \imag\nu_{r})
\ g_{k}^{(0)}(\omega_{o} - \imag\nu_{r+s})
\ g_{m}^{(0)}(\omega_{o} - \imag\nu_{r+s+t})
\nonumber \\
& & + \ n(\epsilon_{k})
\ g_{i}^{(0)}(\omega_{k} + \imag\nu_{s})
\ g_{o}^{(0)}(\omega_{k} + \imag\nu_{r+s})
\ g_{m}^{(0)}(\omega_{k} - \imag\nu_{t})
\nonumber \\
& & \left. + \ n(\epsilon_{m})
\ g_{i}^{(0)}(\omega_{m} + \imag\nu_{s+t})
\ g_{o}^{(0)}(\omega_{m} + \imag\nu_{r+s+t})
\ g_{k}^{(0)}(\omega_{m} + \imag\nu_{t})
\right\} \,. \nonumber
\end{eqnarray}
The $\rho_{rst}$ or the higher order coefficients are of similar
structure.

Recently for the many body system studied here,
a variational approach to the quantum corrections
$\zeta$ of (\ref{Feff}) has been developed in
\cite{rummel:phd,ruc.hoh:var-PF}.
Based on the Euclidean action of the PSPA (\ref{s-PSPA})
a reference action
\begin{equation}\label{s-var}
s_{\Omega}^{q_{0}} =
\frac{\hbar\beta}{|k|} \sum_{r>0} \Lambda_{r} \,|q_{r}|^{2} \,.
\end{equation}
is introduced, where
the RPA frequencies $\varpi_{\mu}^{2}$ appearing in (\ref{lambda})
are replaced by variational parameters $\Omega_{\mu}^{2}$:
\begin{equation}\label{defLambda}
\Lambda_{r}(\beta, q_{0}; \Omega_{\mu}) =
\frac{\prod_{\mu}  (\nu_{r}^{2} + \Omega_{\mu}^{2}(\beta,q_{0}))}
     {\prod_{k>l}' (\nu_{r}^{2} + \omega_{kl}^{2}(q_{0}))}
\end{equation}
The $\Omega_{\mu}^{2}$ are adjusted such that the expectation
value $\langle s_{\textrm{E}} - s_{\Omega}^{q_{0}}
\rangle_{\Omega}^{q_{0}}$ with respect to the $s_{\Omega}^{q_{0}}$
and consequently the effective free energy (\ref{Feff}) is
minimized. Accounting in the action (\ref{defA}) for inharmonic
terms up to fourth order and using the abbreviation
\begin{equation}\label{defPir}
\Pi_{r}(\beta, q_{0}; \Omega_{\mu}) =
\frac{\prod_{\mu} (\nu_{r}^{2} + \varpi_{\mu}^{2}) -
      \prod_{\mu} (\nu_{r}^{2} + \Omega_{\mu}^{2})}
     {\prod_{\mu} (\nu_{r}^{2} + \Omega_{\mu}^{2})}
\end{equation}
within this approximation
the correction factor of the variational approach can be written as
\cite{rummel:phd,ruc.hoh:var-PF}
\begin{equation}\label{defCFKV}
\textrm{ln} \,\zeta^\textrm{(4)}(\beta, q_{0}) =
-\sum_{r>0} \textrm{ln} \Lambda_{r} - \sum_{r>0} \Pi_{r}
- \frac{|k|}{\beta} \sum_{r,s>0} \sigma_{rs-r-s}
\,\frac{1}{\Lambda_{r}} \,\frac{1}{\Lambda_{s}} \,.
\end{equation}
Here, the fourth order coefficient $\sigma_{rstu}$ again
is calculated according to (\ref{Cauchy}).
For convergence of the double sum in the second line of
(\ref{defCFKV}) the behavior of $\sigma_{rstu}$ at large
$r, s, \ldots$ is crucial. By inspection of (\ref{Cauchy}),
using the unperturbed Green's functions (\ref{Green-freq-IPM})
and $\nu_{r} \sim r$, one can convince one-self that
$\sigma_{rstu}$ falls off rapidly enough to assure convergence.

This variational approach \cite{rummel:phd,ruc.hoh:var-PF} to the
partition function of an interacting many body system improves the
accuracy of the PSPA and the ePSPA. Moreover, even in cases where
the static free energy $\mathcal{F}^{\textrm{SPA}}$ develops a
barrier (as a function of $q_{0}$), this novel approach still is
applicable below $T_{0}$ or $T_{0}/2$. Evidently, this variational
approach is the easier to handle the fewer variational parameters
$\Omega_{\mu}$ one needs to introduce. For the present formulation
the latter are unequivocally related to the RPA frequencies
$\varpi_{\mu}$ of the secular equation (\ref{secular}). In
principle, the latter has as many solutions as there are p-h
excitations. However, as already discussed below (\ref{lambda}),
not all of them are equally important for collective motion. In
the ideal case the latter is dominated by only a few or eventually
by just one prominent mode. Then it suffices to concentrate on
these or this latter one(s) \cite{ruc.hoh:var-PF}.

\section{Accounting for residual interactions}
\label{chap-diss}

So far the operator $\hat{H}$ appearing in (\ref{twobodham}) and
(\ref{1bHam}) is identified with that of the independent particle
model (IPM), with the corresponding excitations given by
(\ref{nuclfreq}) and the one body Green's functions by
(\ref{Green-freq-IPM}).
The response function defined by (\ref{defchi}) can be written in the
form $\chi(\omega) = \chi'(\omega) + \imag \chi''(\omega)$.
Calculated from the IPM its imaginary part $\chi''$ consists
of a sum of $\delta$-functions located at {\em discrete} energies.
However, the IPM neglects the
{\em residual two body interaction} between the nucleons,
which describes the incoherent scattering of
particles and holes and consequently couples the $1$p$1$h
excitations of (\ref{1bHam}) to $n$p$n$h
excitations and true compound states. It is this residual
interaction which in the end is responsible for genuine relaxation
processes and, hence, for damping of collective motion. To account
for such effects we will take a pragmatic point of view.
Rather than to embark on the difficult (if at all feasible)
task to account for the residual interaction in
full glory we will simply take over the factorization of the
response function into one body Green's functions $g_{k}(z)$,
which using the corresponding spectral densities $\varrho_{k}(z)$
leads to
\begin{equation}\label{resp-Green}
\chi(z) =
-\hbar \sum_{l,k} F_{lk} F_{kl}
\int_{-\infty}^{\infty} \frac{d\Omega}{2\pi} \ n(\hbar\Omega + \mu)
\left[ g_{k}(z + \Omega) \varrho_{l}(\Omega) +
\varrho_{k}(\Omega) g_{l}(\Omega - z) \right] \,.
\end{equation}
For a more detailed description we refer to \cite{hoh:pr:97}
where also references to earlier work can be found.
Different to the IPM we {\em dress} the Green's functions by
self-energies $\Sigma_{k} = \Sigma_{k}' - \imag\Gamma_{k}/2$
and write
\begin{equation}\label{Green-freq}
g_{k}^{(\Gamma)}(z) =
\frac{1}{\hbar} \,\frac{1}{z - \omega_{k} - \Sigma_{k}(z)/\hbar} \,.
\end{equation}
instead of (\ref{Green-freq-IPM}).
These self-energies are calculated from
\begin{equation}\label{selfen}
\Sigma_{k}(z) =
\int_{-\infty}^{\infty} \frac{d\Omega}{2\pi}
\,\frac{\Gamma_{k}(\Omega)}{z - \Omega}
\end{equation}
with the following {\em phenomenological ansatz} for the widths
\begin{equation}\label{selfimag}
\Gamma_{k}(\omega) = \frac{1}{\Gamma_{0}}
\,\frac{(\hbar\omega - \mu)^{2} + \pi^{2} T^{2}} {1 +
\frac{1}{c^{2}} [(\hbar\omega - \mu)^{2} + \pi^{2} T^{2}]}\,.
\end{equation}
As for the two parameters introduced this way, the values
$\Gamma_{0} \approx 33~\textrm{MeV}$ and $c \approx
20~\textrm{MeV}$ have been used in the past (see e.g.
\cite{siemensetal:84} and \cite{hoh:pr:97}). It can be said that
at zero temperature the widths calculated this way are in good
agreement with empirical data for single particle excitations
\cite{mahauxsartor:91}.
Moreover, at $T = 0$ forms of this type have been used in
\cite{brg.rhm.npa:81} in analyses of experimental results within
optical model type approaches, for more details see Sect.~4.2.3. of
\cite{hoh:pr:97}.

We should like to note that the factorization assumption leading to
(\ref{resp-Green}) is applied to describe the {\em intrinsic}
dynamics but {\em not} that of collective modes.
One may therefore argue that coherent effects are small,
in particular at larger thermal excitations.
After all the underlying SPA is meant to represent the high
temperature limit.

We may proceed now to establish connection to microscopic
transport theories like the one described in \cite{hoh:pr:97}. To
this end it us useful to introduce the {\em collective} response
function defined by
\begin{equation}\label{defchicoll}
\delta\langle\hat{F}\rangle_{\omega} =
-\chi_{\textrm{coll}}(\omega) \, f_{\textrm{ext}}(\omega) \,,
\end{equation}
where $f_{\textrm{ext}}$ stands for an external field coupled to
(\ref{1bHam}) via $f_{\textrm{ext}}(\tau) \hat{F}$. As is well
known (see e.g. \cite{bohra.mottelsonb.2} or \cite{hoh:pr:97}) it
can be obtained from the nucleonic response function of
(\ref{defchi}) through
\begin{equation}\label{chicoll}
\chi_{\textrm{coll}}(\omega) =
\frac{\chi(\omega)}{1 + k \,\chi(\omega)}
\end{equation}
Its poles are given by the solutions of the secular equation
(\ref{secular}). Very similar to the situation discussed in
connection with (\ref{lambda}), not all those poles $\varpi_{\mu}$
contribute equally strongly. In the ideal case {\em one} of them
is shifted down to frequencies which are much smaller than the
corresponding nucleonic excitations
($|\varpi_{\mu}|^{2} \ll \omega_{kl}^{2}$ for stable iso-scalar
modes). Moreover, it may acquire a
considerable part of the overall strength.

In order to take into account dissipative effects we perform two
modifications. As mentioned previously,  the nucleonic response
function $\chi(\omega)$ is calculated from the dressed Green's
functions (\ref{Green-freq}) instead of (\ref{Green-freq-IPM}).
Using the $\Gamma$ of (\ref{selfimag}) the imaginary part
$(\chi^{\Gamma})''(\omega)$ becomes a  continuous function of
$\omega$. As we are interested in a description of {\em slow}
iso-scalar collective motion, as given for instance for nuclear
fission, we will concentrate on the lowest lying (pair of) poles
of the (locally defined) collective response function
(\ref{chicoll}). This truncated form is fitted by the response
function of a damped harmonic oscillator of
inertia $M$, stiffness $C$ and damping $\gamma$
\begin{equation}\label{oscresp2}
\chi_{\textrm{osc}}(\omega) =
\frac{-1}{M\omega^{2} + \imag\gamma\omega - C} =
\frac{1}{|C|}
\,\frac{-1}{(\omega/\varpi)^{2} + 2\imag\eta\omega/\varpi
- \textrm{sgn}C} \,.
\end{equation}
Then, instead of several $\varpi_{\mu}$ only one collective mode
$\varpi$ is left over. The frequency $\varpi$ and the effective
damping coefficient $\eta$ are defined as follows:
\begin{equation}\label{varpieta}
\varpi^{2} = \frac{|C|}{M}
\qquad \textrm{and} \qquad
\eta = \frac{\gamma}{2M\varpi} = \frac{\gamma}{2\sqrt{M|C|}}
\end{equation}
The parameter $\eta$ indicates underdamped ($\eta < 1$) or
overdamped motion ($\eta > 1$). After these modifications, which
may be summarized as
\begin{equation}\label{replacechi}
\chi_{\textrm{coll}}^{\textrm{IPM}}(\omega)
\longrightarrow \chi_{\textrm{coll}}^{\Gamma}(\omega)
\longrightarrow \chi_{\textrm{osc}}(\omega)\,,
\end{equation}
the secular equation (\ref{secular}) for the local collective
modes reduces to the equation
\begin{equation}\label{secular-damp}
\left( \frac{\omega}{\varpi} \right)^{2} +
2\imag\eta \left( \frac{\omega}{\varpi} \right) -
\textrm{sgn}C = 0 \,.
\end{equation}
Its solutions are given by the two frequencies
\begin{equation}\label{coll-damp}
\omega_{\pm} =
\varpi \left( \pm\sqrt{\textrm{sgn}C - \eta^{2}} - \imag\eta \right) \,.
\end{equation}
In contrast to the RPA frequencies $\varpi_{\mu}$ of the IPM, the
$\omega_{\pm}$ have a {\em finite imaginary part}. This implies a
{\em damped} local collective motion due to the influence of the
residual interaction.

At this place it is important to stress that the transport
coefficients $C = C(T,q_{0})$, $\eta = \eta(T,q_{0})$ and $\varpi
= \varpi(T,q_{0})$ or $M = M(T,q_{0})$ and $\gamma =
\gamma(T,q_{0})$ have been derived from a {\em microscopic quantum
theory} starting from the separable two body interaction
(\ref{twobodham}). Like the response function itself, all of them
depend on the temperature $T = 1/\beta$ and the collective
coordinate $q_{0}$. With respect to nuclear collective motion this
dependence may be considered a great advantage over the
Caldeira-Leggett model (CLM) \cite{caa.lea:ap:83}, which is often
used to describe open quantum systems (see e.g. \cite{weissu}; for
a criticism from nuclear physics point of view, see
\cite{hoh:pr:97,rummel:phd}). There, the transport coefficients
are not calculated microscopically. Rather they are introduced as
parameters which are independent of $T$ and $q_{0}$. This latter
feature goes along with the fact that in the CLM the set of
intrinsic degrees of freedom, which acts as the ``heat bath'' for
the collective ones, is modeled by a set of {\em fixed} harmonic
oscillators. The latter do not vary with the collective degree of
freedom, and so does the spectral density. Such an  assumption is
not applicable to  nuclear fission. There the heat bath for
collective motion is given by the nucleonic degrees of freedom
which move in a mean field which depends on the nuclear shape and
thus varies with the collective variable.

\section[Modification of the approximations]
        {Dissipative systems:
         Modification of the approximations to the partition function}
\label{sec-modify}

In this section we study how the inclusion of dissipative effects
modifies the approximations put together in
Sect.~\ref{sec-partfunc} in formal sense. First we look at the PSPA.
After the reduction (\ref{replacechi}) to a single damped mode an
evaluation of
\begin{equation}\label{lambda2}
\lambda_{r} =
1 + k \,\chi(\imag\nu_{r}) =
\frac{1}{1 - k \,\chi_{\textrm{coll}}(\imag\nu_{r})}
\end{equation}
(see (\ref{lambda}) and (\ref{chicoll})) leads to
\begin{equation}\label{lambda-osc}
\lambda_{r} \to \lambda_{r}^{(\Gamma)} =
\frac{\nu_{r}^{2} + \nu_{r}/\tau_{\textrm{kin}} + \textrm{sgn}C \,\varpi^{2}}
     {\nu_{r}^{2} + \nu_{r}/\tau_{\textrm{kin}} + w^{2}} \,.
\end{equation}
Here, the (inverse) time scale
\begin{equation}\label{taukin}
\frac{1}{\tau_{\textrm{kin}}} = \frac{\gamma}{M}
\end{equation}
and the quantity
\begin{equation}\label{freq-nucl}
w^{2}(\beta,q_{0}) = \frac{C - k}{M} =
\textrm{sgn}C \,\varpi^{2}(\beta,q_{0}) + \frac{|k|}{M(\beta,q_{0})} \ge
\textrm{sgn}C \,\varpi^{2}(\beta,q_{0})
\end{equation}
have been introduced.
A calculation of the nucleonic response function
\begin{equation}\label{chinucl-osc}
\chi(\omega) =
\frac{-1}{M\omega^{2} + \imag\gamma\omega - (C - k)}
\end{equation}
from (\ref{oscresp2}) by inversion of the formula (\ref{chicoll})
shows that $w^{2}$ plays the role of the nucleonic frequencies.
In {\em dissipative PSPA} the form of the quantum corrections to the
classical partition function defined in (\ref{Z-athaly})
does not change as compared to the IPM (\ref{defCPSPA}).
Only the $\lambda_{r}$ are replaced by the
$\lambda_{r}^{(\Gamma)}$ of (\ref{lambda-osc}).
The convergence condition (\ref{convcond}) now turns to
$\lambda_{1}^{(\Gamma)}(\beta,q_{0}) > 0$ and explicitly reads
\begin{equation}\label{condT0diss}
\left( \frac{2\pi}{\hbar\beta} \right)^{2} +
\frac{1}{\tau_{\textrm{kin}}} \,\frac{2\pi}{\hbar\beta} +
\textrm{sgn}C \,\varpi^{2} > 0
\end{equation}
In the case of stable modes with $C > 0$ this condition is always
fulfilled. For unstable modes with  $C < 0$, in analogy to
(\ref{T0}) the local crossover temperature $T_{0}(q_{0})$ is given
by the smallest root of the l.h.s. of the inequality
(\ref{condT0diss}). Using the effective damping $\eta$ of
(\ref{varpieta}) this $T_{0}(q_{0})$ becomes
\begin{equation}\label{T0diss}
T_{0}(q_{0}) =
\frac{\hbar\varpi(q_{0})}{2\pi}
\left( \sqrt{1 + \eta^{2}(q_{0})} - \eta(q_{0}) \right) \,.
\end{equation}
Its maximal value $T_{0} = \textrm{Max}_{q_{0}} T_{0}(q_{0})$
defines the global crossover temperature. Like in the CLM for
dissipative quantum systems (see e.g.
\cite{grh.olp.weu:prb:87,hap.tap.bom:rmp:90,weissu}) we find that
damping $\eta > 0$ diminishes this $T_{0}$.

Concerning the anharmonic terms of the Euclidean action there is
an important difference between our approach and the CLM. In the
latter the conservative and the dissipative forces are decoupled
from each other. As one implication, for instance, the inharmonic
terms of the potential $V^{(n)}(q)$ with $n > 2$ are not modified
by dissipative effects. In the case of interacting many body
systems the situation is more complicated: The replacement of the
unperturbed Green's functions (\ref{Green-freq-IPM}) by the
dressed ones (\ref{Green-freq}) not only leads to a finite damping
strength $\eta$ and a change in the frequency of the local
harmonic motion $\omega_{\pm}$. Due to the structure
(\ref{Cauchy}) of the expansion coefficients of the Euclidean
action (\ref{defA}) also the inharmonic terms are influenced by
this replacement, such that $\rho_{rst} \to \rho_{rst}^{(\Gamma)}$
and $\sigma_{rstu} \to \sigma_{rstu}^{(\Gamma)}$. Taking into
account dissipative effects for the quantum corrections of the
ePSPA along the lines of \cite{ruc.anj:epjb:02} one can easily
rederive (\ref{C-ePSPA}), but with $\lambda_{r} \to
\lambda_{r}^{(\Gamma)}$ and
\begin{equation}\label{x-diss}
x \to x^{(\Gamma)} =
\sqrt{\frac{\beta}{4|k|B^{(\Gamma)}}} \ \lambda_{1}^{(\Gamma)} \,,
\end{equation}
where
\begin{equation}\label{B-diss}
B^{(\Gamma)} =
6 \sigma_{1,1,-1,-1}^{(\Gamma)} -
\frac{9 \rho_{-1,-1,2}^{(\Gamma)} \,\rho_{1,1,-2}^{(\Gamma)}}
{\lambda_{2}^{(\Gamma)}} \,.
\end{equation}
Using the variational approach in order to account for dissipative
features the form
\begin{equation}\label{replace_diss}
s_{\Omega}^{q_{0}} =
\frac{\hbar\beta}{|k|} \sum_{r>0}
\Lambda_{r}^{(\Gamma)} \,|q_{r}|^{2}
\end{equation}
with
\begin{equation}\label{defLambda-osc}
\Lambda_{r}^{(\Gamma)}(\beta, q_{0}; \Omega) =
\frac{\nu_{r}^{2} + \nu_{r}/\tau_{\textrm{kin}} + \Omega^{2}}
     {\nu_{r}^{2} + \nu_{r}/\tau_{\textrm{kin}} + w^{2}}
\end{equation}
is used as reference action. The trial action (\ref{replace_diss})
contains {\em only one} variational parameter $\Omega^{2}$ and,
analogously to the case sketched in Sect.~\ref{sec-partfunc},
emerges from the dissipative version of
$s_{\textrm{E}}^{\textrm{PSPA}}$ (see (\ref{s-PSPA}) with
$\lambda_{r}^{(\Gamma)}$ of (\ref{lambda-osc}) instead of
$\lambda_{r}$) by the replacement
$\textrm{sgn} C \,\varpi^{2} \to \Omega^{2}$. Using
(\ref{replace_diss}) and repeating the derivation of
\cite{ruc.hoh:var-PF} for the dissipative case the
quantum correction factor to the classical partition function
in the variational approach turns out to be given by (\ref{defCFKV})
with the replacements $\Lambda_{r} \to \Lambda_{r}^{(\Gamma)}$,
\begin{equation}\label{defPir-osc}
\Pi_{r}
\quad \rightarrow \quad
\Pi_{r}^{(\Gamma)}(\beta, q_{0}; \Omega) =
\frac{\textrm{sgn}C \,\varpi^{2} - \Omega^{2}}
     {\nu_{r}^{2} + \nu_{r}/\tau_{\textrm{kin}} + \Omega^{2}}
\end{equation}
and $\sigma_{rstu} \to  \sigma_{rstu}^{(\Gamma)}$ everywhere:
\begin{eqnarray}\label{C_var_diss}
\textrm{ln} \,\zeta^\textrm{(4)}(\beta, q_{0}) =
- \sum_{r>0} \textrm{ln} \Lambda_{r}^{(\Gamma)}
- \sum_{r>0} \Pi_{r}^{(\Gamma)}
- \frac{|k|}{\beta} \sum_{r,s>0} \sigma_{rs-r-s}^{(\Gamma)}
\,\frac{1}{\Lambda_{r}^{(\Gamma)}} \,\frac{1}{\Lambda_{s}^{(\Gamma)}}
\end{eqnarray}

Please note that this form is very similar to that of the
extension of the Feynman-Kleinert variational approach (FKV)
\cite{fer.klh:pra:86} to open quantum systems (see e.g.
\cite{weissu}). There the quantum partition function of a particle
of mass $M$ in a one-dimensional potential $V(q)$ and an
oscillator heat bath environment has been studied within the CLM.
Using the (inverse) time scale (\ref{taukin}) and the fluctuation
width \cite{weissu}
\begin{equation}\label{a2-diss}
a^{2}(\beta, q_{0}) =
\frac{2}{M\beta} \sum_{r>0}
\frac{1}{\nu_{r}^{2} + \nu_{r}/\tau_{\textrm{kin}}(\nu_{r}) +
\Omega^{2}(\beta,q_{0})}
\end{equation}
to fourth order in $a$ the quantum corrections read \cite{weissu}:
\begin{eqnarray}\label{C_FKV_diss}
\textrm{ln} \,\zeta^{\textrm{FKV}}(\beta,q_{0}) & = &
\sum_{r>0} \textrm{ln} \,\frac{\nu_{r}^{2}}
{\nu_{r}^{2} + \nu_{r}/\tau_{\textrm{kin}}(\nu_{r}) +
\Omega^{2}(\beta,q_{0})} \\
& & -\frac{\beta M}{2} \,[V''(q_{0})/M - \Omega^{2}(\beta, q_{0})]
\,a^{2}(\beta, q_{0}) \nonumber \\
& & -\frac{\beta}{8} \,V^{(4)}(q_{0}) \,(a^{2}(\beta, q_{0}))^{2}
+ \mathcal{O}(a^{6}) \nonumber
\end{eqnarray}
Recalling the definitions of $\Lambda_{r}^{(\Gamma)}$ and
$\Pi_{r}^{(\Gamma)}$ in (\ref{defLambda-osc}) and (\ref{defPir-osc})
respectively, the terms of (\ref{C_var_diss}) and (\ref{C_FKV_diss})
correspond to each other in the order of their appearence.
In (\ref{C_var_diss}) the analog of the squared width
(\ref{a2-diss}) is a sum consisting of terms of the form
$1/\Lambda_{r}^{(\Gamma)}$ (see (\ref{defLambda-osc})),
which converge to $1$ in the limit $r \to \infty$.
As mentioned already before, different to the fourth order term
in (\ref{C_FKV_diss}) for convergence of the double sum
in the last line of (\ref{C_var_diss}) the behavior of
$\sigma_{rs-r-s}^{(\Gamma)}$ for large $r$ and $s$ is crucial.
This difference can finally be traced back to the different path
integral measures in both cases.

\section{Decay of metastable states}
\label{decay}

In \cite{ruc.hoh:pre:01} a formula for the decay rate of damped
interacting many body systems has been derived within the PSPA.
Like that whole formalism it is applicable for $\beta < \beta_{0}$
{\em only}. For $\beta \to \beta_{0}$ the quantum corrections to
the classical rate show an unphysical divergence. From the fact
that the variational approach to the partition function can be
applied also for $\beta > \beta_{0}$ we expect that a rate formula
can be derived, which behaves well in the crossover region and
beyond.

Based on Langer's findings \cite{langer} for the
decay rate $R$ of a metastable system in the literature
(see e.g. \cite{hap.tap.bom:rmp:90} or \cite{weissu})
for high temperatures the formula
\begin{equation}\label{R-T>T0}
R(\beta < \beta_{0}) =
-\frac{2}{\hbar} \ \frac{\beta}{\beta_{0}}
\ \textrm{Im} \,{\mathcal{F}}(\beta) \,,
\end{equation}
is in wide use. With the identifications $1/\beta_{0} = T_{0}$
(form (\ref{T0diss}) evaluated at the barrier $q_{0} = q_{b}$)
and $|\kappa| = -\imag\omega_{+}$ (from (\ref{coll-damp}) with
$\varpi = \varpi_{b}$ and $\textrm{sgn} C = -1$) it becomes identical
to (\ref{rate-langer}). For the low temperature region I.~Affleck has
shown \cite{afi:prl:81} that
\begin{equation}\label{R-T<T0}
R(\beta > \beta_{0}) =
-\frac{2}{\hbar} \ \textrm{Im} \,{\mathcal{F}}(\beta)
\end{equation}
gives the same result as a Boltzmann average over the energy dependent
decay rate $R = -2/\hbar \ \textrm{Im} \,E$. Of course,
the formulas (\ref{R-T>T0}) and (\ref{R-T<T0}) coincide at
$\beta = \beta_{0}$.

In the following we apply (\ref{R-T>T0}) and (\ref{R-T<T0}) to our
modeling of interacting many body systems by the Hamiltonian
(\ref{twobodham}). To this end the integral (\ref{Z-athaly}) for
the partition function and its relation to the free energy of the
total system (\ref{FfromZ}) is used. This integral is dominated by
the free energy $\mathcal{F}^{\textrm{SPA}}(q_{0})$ of the SPA. In
the sequel the latter will be assumed to have just {\em one}
minimum at $q = q_{a}$ and {\em one} barrier at $q = q_{b}$.
(Whenever suitable the indices $a$ and $b$ will be used for
quantities which have to be evaluated at the minimum and the
barrier, respectively.) Moreover, we want to concentrate on
examples where the barrier is sufficiently pronounced. First of
all this implies  that the (temperature-dependent) height
$B^{\textrm{SPA}} = \mathcal{F}_{b}^{\textrm{SPA}} -
\mathcal{F}_{a}^{\textrm{SPA}}$ is sufficiently large, as compared
to temperature,  $\beta B^{\textrm{SPA}}(\beta) \gg 1$. The
minimum and the barrier are assumed to be well separated in
$q_{0}$. In addition we assume that locally they may be approximated
by oscillators with stiffnesses $C^{\textrm{SPA}} =
\partial^{2} \mathcal{F}^{\textrm{SPA}} / \partial q_{0}^{2}$,
with a positive $C^{\textrm{SPA}}_a$
and a negative $C^{\textrm{SPA}}_b$. Any further structure of
$\mathcal{F}^{\textrm{SPA}}$ is neglected. Moreover, the variation
of the quantum corrections $\textrm{ln} \,\zeta(\beta,q_{0})$ with
$q_{0}$ should be sufficiently weak such that the integral
(\ref{Z-athaly}) can be  evaluated by a {\em steepest descent
approximation}. The partition function
\begin{equation}\label{Z-saddle}
\mathcal{Z}(\beta) \approx \mathcal{Z}_{a}(\beta) +
\imag\mathcal{Z}_{b}(\beta)
\end{equation}
then develops an imaginary part which is associated to the
stationary point at the saddle and given by
\begin{equation}\label{Zb} \mathcal{Z}_{b}(\beta) = \frac{1/2}{\sqrt{|k
\,C_{b}^{\textrm{SPA}}(\beta)|}} \ \exp [-\beta
\mathcal{F}_{b}^\textrm{SPA}(\beta)] \ \zeta_{b}(\beta)\,.
\end{equation}
The real part
\begin{equation}\label{Za}
\mathcal{Z}_{a}(\beta) =
\frac{1}{\sqrt{|k| \,C_{a}^{\textrm{SPA}}(\beta)}}
\ \exp [-\beta \mathcal{F}_{a}^\textrm{SPA}(\beta)]
\ \zeta_{a}(\beta)
\end{equation}
comes from the stationary point at the minimum.
To obtain (\ref{Zb}) a special treatment is necessary for the
integration around the barrier top. Following Langer
\cite{langer}, there the contour of integration  must be distorted
into the upper complex half plane. Because of the assumptions
about the barrier height we have $\textrm{Im}\mathcal{Z}(\beta)
\ll \textrm{Re}\mathcal{Z}(\beta)$, implying that the imaginary
part of the free energy can be approximated by
\begin{equation}\label{ImF}
\textrm{Im}\mathcal{F}(\beta) \approx -1/\beta \cdot
\textrm{Im}\mathcal{Z}(\beta) / \textrm{Re}\mathcal{Z}(\beta) \,.
\end{equation}
Using (\ref{R-T>T0}) or (\ref{R-T<T0}) the rate can be separated
into a product of two factors, one representing the classical
limit and the other the quantum corrections:
\begin{equation}\label{rate-gen}
R(\beta) = R_{\textrm{class}}(\beta) \cdot f_{\textrm{qm}}(\beta)
\end{equation}

Inserting the crossover temperature $T_{0} = 1/\beta_{0}$ of
(\ref{T0diss}) and the imaginary part of the free energy (\ref{ImF})
into (\ref{R-T>T0}), for high temperatures the {\em classical part}
reads:
\begin{equation}\label{rate-highT-class}
R_{\textrm{class}}^{\textrm{SPA}}(\beta < \beta_{0}) =
\frac{\varpi_{b}(\beta)}{2\pi}
\left( \sqrt{1 + \eta_{b}^{2}(\beta)} - \eta_{b}(\beta) \right)
\sqrt{\frac{C_{a}^\textrm{SPA}(\beta)}{C_{b}^\textrm{SPA}(\beta)}}
\,\exp [-\beta B^\textrm{SPA}(\beta)]
\end{equation}
This result has already been found in \cite{ruc.hoh:pre:01}.
There it also had been rewritten in a more intuitive way by making use
of (\ref{varpieta}) for the stiffnesses:
\begin{equation}\label{inertias}
\frac{\varpi_{b}}{2\pi}
\,\sqrt{\frac{C_{a}^{\textrm{SPA}}}{C_{b}^{\textrm{SPA}}}} =
\frac{\varpi_{a}}{2\pi} \,\sqrt{\frac{M_{a}}{M_{b}}}
\end{equation}
In this way in (\ref{rate-highT-class}) besides the Arrhenius
factor $\exp [-\beta B^\textrm{SPA}]$ the attempt frequency
$\varpi_{a}$ appears, with which the system tries to overcome the
barrier. Notice please that in this rate formula the factor in
brackets is equivalent to Kramers' famous correction factor
\cite{krh:ph:40}. It indicates that the classical contribution to
the rate is {\em reduced by damping}. Different to the rate
formulas derived in the CLM in (\ref{inertias}) the ratio of the
inertias at the barrier and the minimum appears
\cite{ruc.hoh:pre:01}. It should be mentioned, however, that this
additional factor is not seen in Langer's original work although
his derivation started from a Fokker-Planck equation where the
inertia is allowed to depend on the collective variable. To
clarify this point one may eventually have to modify the way the
collective degree of freedom is introduced in the common SPA and
PSPA formulations of the HST on which also our approach bases
heavily.

For low temperatures application of (\ref{R-T<T0}) leads to
\begin{equation}\label{rate-lowT-class}
R_{\textrm{class}}^{\textrm{SPA}}(\beta > \beta_{0}) =
\frac{1}{\hbar\beta}
\,\sqrt{\frac{C_{a}^\textrm{SPA}(\beta)}{C_{b}^\textrm{SPA}(\beta)}}
\,\exp [-\beta B^\textrm{SPA}(\beta)] \,.
\end{equation}
Note that this region has not been accessible with the PSPA in
\cite{ruc.hoh:pre:01}. The formulas (\ref{rate-highT-class}) and
(\ref{rate-lowT-class}) describe the contribution of {\em thermal
activation} to the decay rate.

Finally, we turn to the {\em quantum correction} factor
$f_{\textrm{qm}}(\beta)$. Both for (\ref{rate-highT-class}) as
well as for (\ref{rate-lowT-class}) it has the same form,
\begin{equation}\label{quantcorr}
f_{\textrm{qm}}(\beta) =
\frac{\zeta_{b}(\beta)}{\zeta_{a}(\beta)}
\end{equation}
and {\em increases} the rate. In \cite{ruc.hoh:pre:01} it has
already been pointed out for the PSPA that using (\ref{defCPSPA}) with
(\ref{lambda-osc}) the factor (\ref{quantcorr}) has better convergence
properties than the ad-hoc generalization of the CLM to
coordinate-dependent damping. In the following for the evaluation of
$\zeta$ and $f_{\textrm{qm}}$ we will limit ourselves to the
variational approach.

\section{An estimate for the accuracy of the rate formula}
\label{reliability}

In Sect.~\ref{decay} the framework for the calculation of the decay
rate of metastable self-bound Fermi systems such as heavy atomic
nuclei has been developed. To test the accuracy of the formula
(\ref{rate-gen}) with (\ref{rate-highT-class}) or
(\ref{rate-lowT-class}) and (\ref{quantcorr}) it would be desirable
to have at one's disposal a model for {\em metastable} many body systems,
for which an exact result can be given even in the case of
{\em finite damping}.
Unfortunately, to the best of our knowledge, such a model does
not exist. One might think to take the Lipkin-Meshkov-Glick model
(LMGM) \cite{lih.men.gla:np:65} in the version of
\cite{ruc.hoh:var-PF}. Indeed, in this model the free energy
$\mathcal{F}^{\textrm{SPA}}(q_{0})$ develops a barrier below a
critical temperature. This barrier separates two minima in a
symmetric, bound system. Therefore it is not possible to introduce
a decay rate in proper sense. Also based on the LMGM in
\cite{arve:prc:87} an exactly solvable model for
many body systems has been developed, which allows to study the
fission process. Unfortunately, both models do not allow for exact
solutions at finite damping.

Because of this problem we will follow a different strategy and
look at the much simpler one-dimensional case of a particle of
mass $M$ moving in the cubic potential
\begin{equation}\label{potV}
V(q) =
\frac{M}{2} \,\varpi^{2} \,q^{2} \left( 1 - \frac{q}{Q} \right)
\end{equation}
with minimum at $q_{a} = 0$, barrier at $q_{b} = 2Q/3$ and with a
barrier height $B = V_{b} - V_{a} = 2 \,M\varpi^{2} \,Q^{2}/27$,
although our approach is suitable for many body systems.
For this situation the quantum decay rate has been calculated
{\em exactly} with help of the CLM for all temperatures
$T = 0 \ldots \infty$ and various damping strengths
\cite{caa.lea:ap:83,larkin,grabert,grh.olp.weu:prb:87}. In
addition, in the FKV there exists an analog of the variational
approach used in the present paper, namely for a particle moving
in a one-dimensional potential.

Unfortunately, in its original form \cite{fer.klh:pra:86} the FKV
is not directly applicable to the potential (\ref{potV}). The
reason is that in the expansion (\ref{C_FKV_diss}) only
derivatives of $V(q_{0})$ of {\em even} order contribute.
Consequently, like for the harmonic oscillator, for a cubic
potential the FKV reduces to a Gaussian approximation and
inharmonic terms are not seen. To cure this problem, in
\cite{klh.mui:ijmpa:96} besides the trial frequency $\Omega$ a
second variational parameter has been introduced, which controls
the minimum of the local oscillator. Here, we follow a simpler
strategy and consider the {\em stable} fourth order potential
\begin{equation}\label{potW}
W(q, n, C') = C' \,\frac{M}{2} \,\varpi^{2} \,q^{2}
\left( 1 - \frac{q}{Q} \right) \left( 1 - \frac{q}{n Q} \right)
\,,
\end{equation}
which by adjusting $n > 1$ and $C'$ can be fixed such that it
reproduces the barrier region of (\ref{potV}) to arbitrarily high
accuracy. By application of the formulas (\ref{R-T>T0}) and
(\ref{R-T<T0}), for which the free energy has to be evaluated for
the particle of mass $M$ in the potential (\ref{potW}), one gets
(see e.g. \cite{hap.tap.bom:rmp:90,weissu})
\begin{eqnarray}
R_{\textrm{class}}(\beta < \beta_{0}) & = &
\frac{\sqrt{|W_{b}''|/M}}{2\pi} \left( \sqrt{1 + \eta^{2}} - \eta
\right) \sqrt{\frac{W_{a}''}{|W_{b}''|}}
\,\exp (-\beta B) \label{rate-V-lowT} \\
R_{\textrm{class}}(\beta > \beta_{0}) & = & \frac{1}{\hbar\beta}
\sqrt{\frac{W_{a}''}{|W_{b}''|}} \exp (-\beta B)
\label{rate-V-highT}
\end{eqnarray}
instead of (\ref{rate-highT-class})--(\ref{rate-lowT-class}).
Using (\ref{C_FKV_diss}) for the calculation of the correction
factor (\ref{quantcorr}) one can get an {\em estimate} on the
accuracy of the formulas derived in Sect.~\ref{decay}.

\begin{figure}[htb] \begin{center}
\epsfig{file=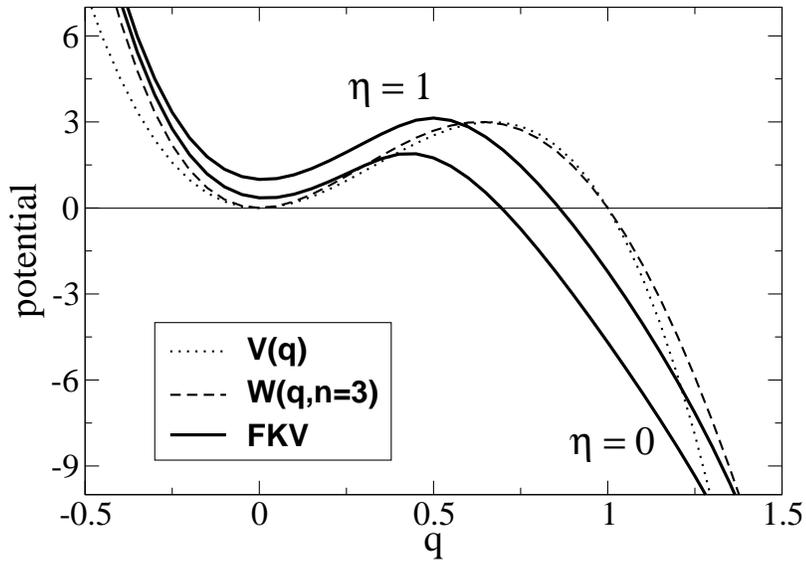, height=120mm, angle=-90}
\caption{\label{fig-VWWeff} The metastable potentials in one
dimension, for a barrier height of $B=3\hbar \varpi$. The
coordinate $q$ is given in units of $Q$ and energies are scaled to
$\hbar\varpi$. The effective potential $W^{\textrm{FKV}}(\beta,q)$
is shown for $\beta = 2\beta_{0}$ and two effective damping
strengths $\eta = 0$ and $\eta = 1$.}
\end{center} \end{figure}
In Fig.~\ref{fig-VWWeff} we show the barrier region of the
potentials $V(q)$ and $W(q, n=3)$ with barrier height $B =
3\hbar\varpi$. The global minimum of the potential $W(q, n=3)$ is
found at $q_{+} \approx 2.4 Q$ (not shown in the figure) and is of
value $-W(q_{+}) \approx 42\hbar\varpi \gg B$. Because of the
large depth of this minimum a thermally activated flux originating
from this minimum is strongly suppressed. Also shown in
Fig.~\ref{fig-VWWeff} is the ``effective classical potential''
\cite{fer.klh:pra:86}
\begin{equation}\label{WFKV}
W^{\textrm{FKV}}(\beta,q_{0}) = W(q_{0}) -
\frac{1}{\beta} \ \textrm{ln} \,\zeta^{\textrm{FKV}}(\beta,q_{0})
\end{equation}
for two effective damping strengths and a given temperature.
It is the analog of (\ref{Feff}) evaluated in the variational
approach and contains quantum corrections $\zeta^{\textrm{FKV}}$,
which are taken from (\ref{C_FKV_diss}) with $V$ replaced by $W$.
As compared to $W(q_{0})$ the effective classical potential
$W^{\textrm{FKV}}(\beta,q_{0})$ has a smaller barrier and slightly
smaller stiffnesses at $q_{a}$ and $q_{b}$.

\begin{figure}[htb] \begin{center}
\epsfig{file=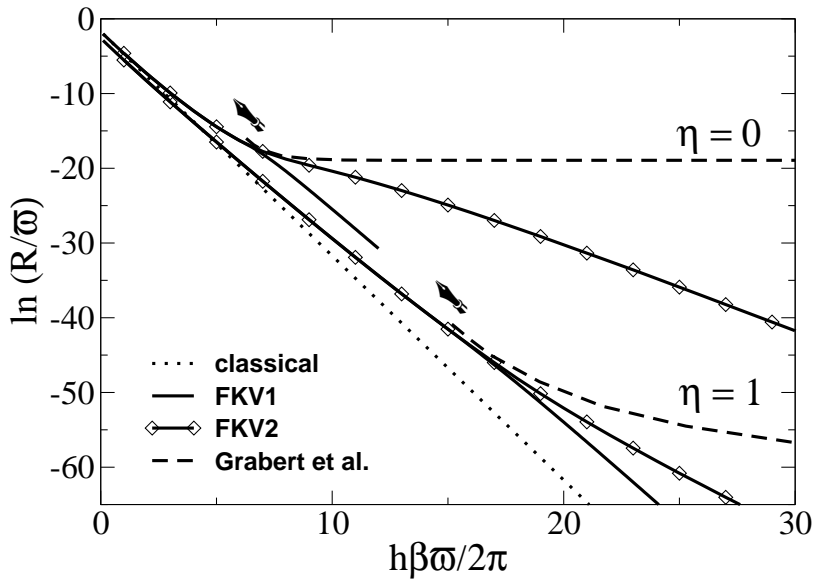, height=120mm, angle=-90}
\caption{\label{fig-rateW} A comparison of different approaches to
the decay rate for a particle in a one-dimensional cubic potential
with barrier, for zero ($\eta = 0$) and critical ($\eta = 1$)
damping. The arrows point to the $\beta_{0} = 1/T_{0}$ where the
PSPA breaks down. For more details see text.}
\end{center} \end{figure}
In Fig.~\ref{fig-rateW} we show the $\beta$-dependence of the rate
formulas in logarithmic representation, for two effective damping
strengths. The results of two different calculations are
presented. The one denoted by ``FKV1'' (fully drawn line) is
evaluated by using formulas (\ref{rate-V-lowT}) and
(\ref{rate-V-highT}), with the quantum corrections
$f_{\textrm{qm}}$ taken from (\ref{quantcorr}) and
(\ref{C_FKV_diss}). For the second calculation
(``FKV2'', fully drawn line with symbols) the saddle
point approximation is applied directly to the effective classical
potential (\ref{WFKV}). The {\em quantal} rate is calculated from
the formulas
\begin{eqnarray}
R(\beta < \beta_{0}) & \!=\! &
\frac{\sqrt{|(W_{b}^{\textrm{FKV}})''|/M}}{2\pi} \left( \sqrt{1 +
\eta^{2}} - \eta \right)
\sqrt{\frac{(W_{a}^{\textrm{FKV}})''}{|(W_{b}^{\textrm{FKV}})''|}}
\,\exp (-\beta B^{\textrm{FKV}}) \quad
\label{rate-W-lowT} \\
R(\beta > \beta_{0}) & = & \frac{1}{\hbar\beta}
\,\sqrt{\frac{(W_{a}^{\textrm{FKV}})''}{|(W_{b}^{\textrm{FKV}})''|}}
\,\exp (-\beta B^{\textrm{FKV}}) \,, \label{rate-W-highT}
\end{eqnarray}
where $B^{\textrm{FKV}} = W_{b}^{\textrm{FKV}} -
W_{a}^{\textrm{FKV}}$ is the (temperature-dependent) effective
barrier height within the FKV, which is smaller than $B$, as can
be seen from Fig.~\ref{fig-VWWeff}. Please note that FKV2 is a
less drastic approximation to the $q_{0}$-integral for
$\mathcal{Z}$ than FKV1. In addition we show in
Fig.~\ref{fig-rateW} the classical rate as a dotted line, for which
$f_{\textrm{qm}}^{\textrm{SPA}} \equiv 1$. The analog of the PSPA
rate for a particle moving in a one-dimensional potential takes
into account quantum effects via a Gaussian approximation. In
order to keep the figure transparent these curves are not shown
here. The inverse crossover temperature $\beta_{0} = 1/T_{0}$,
however, where the PSPA formalism breaks down due to a divergence
of $f_{\textrm{qm}}^{\textrm{PSPA}}$, is marked by arrows. For low
temperatures $\beta > \beta_{0}$ also the result of H.~Grabert et
al. \cite{grabert,grh.olp.weu:prb:87} is plotted (dashed line).
Taking from Tab.~I and II of \cite{grh.olp.weu:prb:87}
the relevant quantities, these results
were obtained from an exact numerical treatment of the dynamical
``bounce solution'' $q_{\textrm{B}}(\tau)$.

The classical decay rate shows the purely exponential behavior
known from Arrhenius' law. The rates derived from FKV1 and FKV2
coincide with the classical result at high temperatures (small
$\beta$) but show enhancement at larger $\beta$. In the crossover
region $\beta \approx \beta_{0}$ these approximations smoothly
connect the classical rate with the result of Grabert et al. for
$\beta > \beta_{0}$. For very low temperatures ($\beta \gg
\beta_{0}$), however, FKV1 and FKV2 deliver bad results, too.
Instead of converging to a finite limit, both versions
qualitatively behave more like the classical rate and fall off
exponentially. This can be understood as follows: Also the FKV is
basically a {\em static} approximation that incorporates
inharmonic terms by smearing out the potential locally, see
Fig.~\ref{fig-VWWeff}. This works well as long as the quantum
fluctuations are not too large, as given for moderate to high
temperatures. At very low temperatures, however, there is no way
around accounting for the full nonlinearity of the {\em dynamical}
bounce solution $q_{B}(\tau)$, which minimizes the Euclidean
action for $\beta > \beta_{0}$ and therefore gives the leading
contribution to the decay rate. Due to this limitation of the
variational approach, it would simply be asking too much to expect
good agreement with the fully quantum rate even at extremely small
temperatures. As the FKV2 contains more information about the
quantum effects than the FKV1 it underestimates the rate less
dramatically at low temperatures. Especially in the region $\beta
\gtrsim \beta_{0}$ FKV2 still delivers acceptable results where
FKV1 is no longer reliable.

\end{document}